\documentclass[preprintnumbers,amsmath,amssymbm,prd]{revtex4}
\usepackage{epsfig}
\usepackage{graphicx}
\usepackage{amssymb}

\begin{document}

\title{Upper bounds on the force function in spatially regular self-gravitating matter configurations}
\author{Shahar Hod}
\affiliation{The Ruppin Academic Center, Emeq Hefer 40250, Israel}
\date{\today}

\begin{abstract}

\ \ \ We use the non-linearly coupled Einstein-matter field equations to prove four theorems that bound from above 
the dimensionless force function ${\cal F}=4\pi r^2\cdot p(r)$ in spatially regular curved spacetimes of 
spherically symmetric self-gravitating matter configurations [here $p(r)$ is the radially-dependent 
pressure inside the spatially regular matter configurations]. 
In particular, for generic (not necessarily isotropic) matter configurations it is proved that: 
(i) ${\cal F}\leq 2$ for matter fields that satisfy the dominant energy condition, and 
(ii) ${\cal F}\leq 1$ for matter fields with a non-positive energy-momentum trace. 
In addition, for self-gravitating isotropic matter configurations we derive the stronger upper bounds: 
(iii) ${\cal F}\leq 1$ for matter fields that satisfy the dominant energy condition, and 
(iv) ${\cal F}\leq 1/2$ for matter fields with a non-positive energy-momentum trace. 
Our analytically derived results are in accord with the spirit of the maximum force conjecture in general relativity. 
\end{abstract}
\bigskip
\maketitle

\section{Introduction}

The intriguing assertion made by Gibbons \cite{Gib} and Schiller \cite{Sch1,Sch2} that, 
within the framework of general relativity, forces are bounded from above 
by a relation of the form
\begin{equation}\label{Eq1}
{\cal F}\leq\eta\cdot{{c^4}\over{G}}\
\end{equation}  
with $\eta=O(1)$ \cite{Noteeta} is known as the maximum force conjecture (see also \cite{Ong,Dand,Hodstb}). 
Intriguingly, arguments have been given \cite{Sch3} that the Einstein 
field equations can be deduced from the maximum force relation (\ref{Eq1}). 

Currently, there is no general proof in the physics literature 
of the validity of the maximum force conjecture (\ref{Eq1}) in general relativity. 
The main goal of the present compact paper is to prove, using {\it analytical} techniques, that the 
radially-dependent force function \cite{NoteUnits}
\begin{equation}\label{Eq2}
{\cal F}\equiv 4\pi r^2\cdot p(r)\
\end{equation}
in curved spacetimes of spherically symmetric self-gravitating matter configurations 
is bounded from above by a relation of the form (\ref{Eq1}) 
[here $p(r)$ is the radial pressure inside the self-gravitating matter configurations]. 

In particular, motivated by the physically intriguing conjecture made in \cite{Gib,Sch1,Sch2} 
we shall explicitly prove below that, for generic (that is, not necessarily isotropic) self-gravitating 
matter configurations in curved spacetimes, the non-linearly coupled Einstein-matter field equations set 
an upper bound of the form (\ref{Eq1}) on the dimensionless force function (\ref{Eq2}) with: 
(i) $\eta=2$ for matter configurations that satisfy the dominant energy condition, and
(ii) $\eta=1$ for field configurations with a non-positive energy-momentum trace \cite{Bond1,Beklt,HodT}. 
For self-gravitating isotropic field configurations in curved spacetimes we shall derive the stronger upper bounds: 
(iii) $\eta=1$ for matter configurations that satisfy the dominant energy condition \cite{Noteoth}, and 
(iv) $\eta=1/2$ for field configurations with a non-positive energy-momentum trace \cite{Bond1,Beklt,HodT}.  

\section{Description of the system}

We shall analyze the physical and mathematical properties of self-gravitating matter configurations 
whose spatially regular spacetimes are described, using the Schwarzschild spacetime 
coordinates $(t,r,\theta,\phi)$, by the spherically symmetric curved line element
\cite{Chan,ShTe,Hodt1,Hodt2}
\begin{equation}\label{Eq3}
ds^2=-e^{-2\delta}\mu dt^2 +\mu^{-1}dr^2+r^2(d\theta^2 +\sin^2\theta d\phi^2)\  .
\end{equation}

The radially-dependent dimensionless metric functions $\{\mu(r),\delta(r)\}$ of the 
curved spacetime are determined by the non-linearly coupled Einstein-matter field 
equations. In particular, using the line element (\ref{Eq3}) and the 
relations \cite{Bond1}
\begin{equation}\label{Eq4}
\rho\equiv-T^{t}_{t}\ \ \ \ ,\ \ \ \ p\equiv T^{r}_{r}\ \ \ \ , \ \ \ \ p_T\equiv T^{\theta}_{\theta}=T^{\phi}_{\phi}\
\end{equation}
for the energy density, the radial pressure, and the tangential pressure of 
the self-gravitating matter configurations, 
one can express the Einstein-matter equations $G^{\mu}_{\nu}=8\pi T^{\mu}_{\nu}$ in the differential forms 
\cite{Chan,BekMay,Hodt1}
\begin{equation}\label{Eq5}
{{d\mu}\over{dr}}=-8\pi r\rho+{{1-\mu}\over{r}}\
\end{equation}
and
\begin{equation}\label{Eq6}
{{d\delta}\over{dr}}=-{{4\pi r(\rho +p)}\over{\mu}}\  .
\end{equation}

The metric functions of spatially regular asymptotically flat spacetimes are characterized by the 
boundary conditions \cite{BekMay,Hodt1}
\begin{equation}\label{Eq7}
\mu(r\to 0)\to1\ \ \ \ ; \ \ \ \ \mu(r\to\infty)\to 1\
\end{equation}
and
\begin{equation}\label{Eq8}
\delta(r\to0)<\infty\ \ \ \ ; \ \ \ \ \delta(r\to\infty)\to 0\ .
\end{equation}
In addition, self-gravitating horizonless matter configurations are characterized by the inequality 
\begin{equation}\label{Eq9}
\mu(r)>0\ \ \ \ \text{for}\ \ \ \ r\in[0,\infty]\  .
\end{equation}

From the Einstein-matter differential equation (\ref{Eq5}) one obtains the 
functional relation \cite{BekMay,Hodt1}
\begin{equation}\label{Eq10}
\mu(r)=1-{{2m(r)}\over{r}}\
\end{equation}
for the radially-dependent dimensionless metric function, 
where the integral relation  
\begin{equation}\label{Eq11}
m(r)=\int_{0}^{r} 4\pi x^2\rho(x)dx\
\end{equation}
determines the gravitational mass which is contained within a sphere of radius $r$ \cite{BekMay,Hodt1}. 

The self-gravitating matter configurations that we analyze are assumed to be characterized by 
the dominant energy condition \cite{BekMay},  
\begin{equation}\label{Eq12}
0\leq |p|,|p_{\text{T}}|\leq\rho\  ,
\end{equation}
which, taking cognizance of the relations (\ref{Eq10}) and (\ref{Eq11}), implies 
the characteristic dimensionless inequality 
\begin{equation}\label{Eq13}
\mu(r)\leq1\ \ \ \ \text{for}\ \ \ \ r\in[0,\infty]\  .
\end{equation}

\section{Maximum force theorems for generic self-gravitating matter configurations in spatially regular curved spacetimes}

In the present section we shall derive, using analytical techniques, two upper bounds 
on the dimensionless force function (\ref{Eq2}) that characterizes generic (that is, not necessarily isotropic) 
self-gravitating matter configurations in 
curved spacetimes. The strengths of the upper bounds that we shall derive below would depend on the 
energy conditions which are assumed to be respected by the self-gravitating matter fields. 

We first point out that Eqs. (\ref{Eq7}), (\ref{Eq10}), and (\ref{Eq11}) imply that 
spatially regular matter configurations are characterized by the boundary relations
\begin{equation}\label{Eq14}
r^2\cdot \rho(r)\to 0\ \ \ \ \text{for}\ \ \ \ r\to0\ 
\end{equation}
and
\begin{equation}\label{Eq15}
r^2\cdot \rho(r)\to 0\ \ \ \ \text{for}\ \ \ \ r\to\infty\  .
\end{equation}
Using the dominant energy condition (\ref{Eq12}), one deduces from (\ref{Eq14}) and (\ref{Eq15}) that 
the dimensionless force function (\ref{Eq2}) of spatially regular self-gravitating 
matter configurations is characterized by the simple boundary relations
\begin{equation}\label{Eq16}
{\cal F}(r)\to 0\ \ \ \ \ \text{for}\ \ \ \ \ r\to 0\
\end{equation}
and
\begin{equation}\label{Eq17}
{\cal F}(r)\to 0\ \ \ \ \ \text{for}\ \ \ \ \ r\to\infty\  .
\end{equation}

Interestingly, the characteristic functional behaviors (\ref{Eq16}) and (\ref{Eq17}) 
imply that the radially-dependent force function (\ref{Eq2}) must have an extremum point 
with the property
\begin{equation}\label{Eq18}
{{d{\cal F}}\over{dr}}=0\ \ \ \ \ \text{for}\ \ \ \ \ r=r_{\text{max}}\  .
\end{equation}
Taking cognizance of the Einstein-matter differential equations (\ref{Eq5}) and (\ref{Eq6}) and using the 
conservation equation
\begin{equation}\label{Eq19}
T^{\mu}_{r ;\mu}=0\  ,
\end{equation}
one obtains the gradient relation
\begin{equation}\label{Eq20}
{{d{\cal F}}\over{dr}}={{2\pi r}\over{\mu}}\Big[(3\mu-1-2{\cal F})(\rho+p)+2\mu T-4\mu p\Big]\
%{{d{\cal F}}\over{dr}}={{2\pi r}\over{\mu}}\Big[(3\mu-1-8\pi r^2p)(\rho+p)+2\mu(-\rho-p+2p_{T})\Big]\
\end{equation}
for the dimensionless force function (\ref{Eq2}) of the self-gravitating matter configurations, 
where 
\begin{equation}\label{Eq21}
T=-\rho+p+2p_{T}\
\end{equation}
is the trace of the energy-momentum tensor that characterizes the self-gravitating matter fields.  

\subsection{Upper bound on the dimensionless force function in self-gravitating matter configurations 
that respect the dominant energy condition}

For matter configurations that satisfy the dominant energy condition (\ref{Eq12}), 
one finds from Eqs. (\ref{Eq18}), (\ref{Eq20}), and (\ref{Eq21}) the relations
\begin{equation}\label{Eq22}
0=(\rho+p)\cdot(\mu-1-2{\cal F})+4\mu p_{\text{T}}\leq 
\rho\cdot(5\mu-1-2{\cal F})+p\cdot(\mu-1-2{\cal F})\ \ \ \ \ \text{for}\ \ \ \ \ r=r_{\text{max}}\
\end{equation}
at the extremum point of the dimensionless force function (\ref{Eq2}). 
Note that Eq. (\ref{Eq13}) implies the inequality $p\cdot(\mu-1-2{\cal F})\leq0$ 
for $r=r_{\text{max}}$ \cite{Noteppd,Notefp}, which yields the inequality [see Eq. (\ref{Eq22})]
\begin{equation}\label{Eq23} 
\rho\cdot(5\mu-1-2{\cal F})\geq0\ \ \ \ \ \text{for}\ \ \ \ \ r=r_{\text{max}}\  ,
\end{equation}
or equivalently [see Eq. (\ref{Eq13})] \cite{Noterho}
\begin{equation}\label{Eq24}
{\cal F}_{\text{max}}\leq{5\over2}\cdot\mu(r_{\text{max}})-{1\over2}\leq2\  .
\end{equation}

\subsection{Upper bound on the force function for self-gravitating matter configurations with a 
non-positive energy-momentum trace}

Following \cite{Bond1,Beklt,HodT}, in the present subsection we shall consider self-gravitating matter configurations 
whose energy-momentum tensors have non-positive traces,
\begin{equation}\label{Eq25}
T\leq0\  .
\end{equation}
In particular, we shall explicitly prove that a bound on the force function (\ref{Eq2}), 
which is {\it stronger} than (\ref{Eq24}), can be derived for self-gravitating field configurations 
which are characterized by the non-positive trace condition (\ref{Eq25}) \cite{Bond1,Beklt,HodT}. 

Taking cognizance of Eqs. (\ref{Eq18}), (\ref{Eq20}), and (\ref{Eq25}) one finds 
the compact inequality \cite{Noteppd}
\begin{equation}\label{Eq26}
(\rho+p)\cdot(3\mu-1-2{\cal F})\geq0\ \ \ \ \ \text{for}\ \ \ \ \ r=r_{\text{max}}\
\end{equation}
at the extremum point of the dimensionless force function. 

From Eqs. (\ref{Eq12}), (\ref{Eq13}), and (\ref{Eq26}) one immediately obtains the upper bound 
\begin{equation}\label{Eq27}
{\cal F}_{\text{max}}\leq{3\over2}\cdot\mu(r_{\text{max}})-{1\over2}\leq1\
\end{equation}
on the radially-dependent force function (\ref{Eq2}) for self-gravitating matter configurations with a 
non-positive energy-momentum trace \cite{Bond1,Beklt,HodT}.  

\section{Maximum force theorems for isotropic matter configurations in spatially regular curved spacetimes}

In the present section we shall explicitly prove that, for self-gravitating 
{\it isotropic} matter configurations in curved spacetimes, 
one can derive upper bounds on the dimensionless force function (\ref{Eq2}) 
which are {\it stronger} than the generic bounds (\ref{Eq24}) and (\ref{Eq27}) \cite{Notegen}. 

\subsection{Upper bound on the dimensionless force function for isotropic self-gravitating matter configurations 
that satisfy the dominant energy condition}

For isotropic matter configurations, which are characterized by the simple property \cite{Noteoth}
\begin{equation}\label{Eq28}
p_{T}=p\  ,
\end{equation}
one obtains from Eqs. (\ref{Eq18}), (\ref{Eq20}), (\ref{Eq21}), and (\ref{Eq28}) 
the compact functional relation 
\begin{equation}\label{Eq29}
\rho\cdot(\mu-1-2{\cal F})+p\cdot(5\mu-1-2{\cal F})=0\ \ \ \ \ \text{for}\ \ \ \ \ r=r_{\text{max}}\
\end{equation}
at the extremum point of the dimensionless force function (\ref{Eq2}). 

From Eq. (\ref{Eq29}) and the dominant energy condition (\ref{Eq12}) one finds the characteristic 
inequality 
\begin{equation}\label{Eq30}
\Big[{{p}\over{\rho}}\Big]_{r=r_{\text{max}}}={{1-\mu(r_{\text{max}})+2{\cal F}_{\text{max}}}
\over{5\mu(r_{\text{max}})-1-2{\cal F}_{\text{max}}}}\leq1\  .
\end{equation} 
Note that Eqs. (\ref{Eq12}) and (\ref{Eq13}) imply the relation $\rho\cdot(\mu-1-2{\cal F})\leq0$ 
for $r=r_{\text{max}}$ \cite{Notefp,Noterho}, 
which yields the inequality $p\cdot(5\mu-1-2{\cal F})\geq0$ for $r=r_{\text{max}}$ [see Eq. (\ref{Eq29})], 
or equivalently \cite{Noteppd}
\begin{equation}\label{Eq31}
5\mu-1-2{\cal F}\geq0\ \ \ \ \ \text{for}\ \ \ \ \ r=r_{\text{max}}\  .
\end{equation}

Substituting the inequality (\ref{Eq31}) into Eq. (\ref{Eq30}) one finds the upper bound [see Eq. (\ref{Eq13})] 
\begin{equation}\label{Eq32}
{\cal F}_{\text{max}}\leq{3\over2}\cdot\mu(r_{\text{max}})-{1\over2}\leq1\
\end{equation}
on the dimensionless force function in spatially regular spacetimes of 
isotropic self-gravitating matter configurations. 

\subsection{Upper bound on the force function for isotropic self-gravitating matter configurations with a 
non-positive energy-momentum trace}

In the present subsection we shall prove that a bound on the dimensionless force function (\ref{Eq2}), 
which is stronger than the bound (\ref{Eq27}), can be derived for isotropic self-gravitating matter configurations 
whose energy-momentum tensors have non-negative traces \cite{Bond1,Beklt,HodT}. 
From of Eqs. (\ref{Eq21}), (\ref{Eq25}), and (\ref{Eq28}) 
one finds that these matter models are characterized by the inequality
\begin{equation}\label{Eq33}
p\leq {1\over3}\rho\  .
\end{equation}

Taking cognizance of Eqs. (\ref{Eq18}), (\ref{Eq20}), (\ref{Eq21}), (\ref{Eq28}), and (\ref{Eq33}) 
one obtains the compact functional relation 
\begin{equation}\label{Eq34}
\Big[{{p}\over{\rho}}\Big]_{r=r_{\text{max}}}={{1-\mu(r_{\text{max}})+2{\cal F}_{\text{max}}}
\over{5\mu(r_{\text{max}})-1-2{\cal F}_{\text{max}}}}\leq{1\over3}\
\end{equation}
at the extremum point of the dimensionless force function (\ref{Eq2}). 
The inequality (\ref{Eq34}) yields the upper bound [see Eqs. (\ref{Eq13}) and (\ref{Eq31})]
\begin{equation}\label{Eq35}
{\cal F}_{\text{max}}\leq\mu(r_{\text{max}})-{1\over2}\leq{1\over2}\
\end{equation}
on the dimensionless force function of isotropic self-gravitating matter configurations with 
a non-positive energy-momentum trace \cite{Bond1,Beklt,HodT}.  

\section{Summary}

Motivated by the physically interesting conjecture made in \cite{Gib,Sch1,Sch2} (see also \cite{Ong,Dand,Hodstb}) 
that forces in general relativity are bounded by a mathematically compact 
relation of the form (\ref{Eq1}) we have studied, using analytical techniques, the radial functional behavior of the 
dimensionless force function (\ref{Eq2}) that characterizes spherically symmetric 
self-gravitating matter configurations in curved spacetimes.  

In particular, we have explicitly proved that the non-linearly coupled Einstein-matter field equations set 
the upper bounds 
\begin{eqnarray}\label{Eq36}
{\cal F}\leq
\begin{cases}
2 & \ \ \ \text{for generic field configurations that satisfy the dominant energy condition}\ \\
1 & \ \ \ \text{for generic field configurations with $T\leq0$}\ \\
1 & \ \ \ \text{for isotropic field configurations that satisfy the dominant energy condition}\ \\
{1\over2} & \ \ \ \text{for isotropic field configurations with $T\leq0$}\ 
\end{cases}
\end{eqnarray}
on the dimensionless force function in curved spacetimes of self-gravitating matter configurations. 

Finally, we would like to emphasize that our analytically derived results 
support the spirit of the maximum force conjecture made in the physically important works \cite{Gib,Sch1,Sch2} 
(see also \cite{Ong,Dand,Hodstb}). 
In particular, we have explicitly proved that, 
under physically plausible conditions, forces in general relativity are bounded from above by a 
functional relation of the form (\ref{Eq1}) with $\eta=O(1)$. 

\bigskip
\noindent {\bf ACKNOWLEDGMENTS}
%\bigskip

This research is supported by the Carmel Science Foundation. I would
like to thank Yael Oren, Arbel M. Ongo, Ayelet B. Lata, and Alona B.
Tea for stimulating discussions.


\begin{thebibliography}{99}
\bibitem{Noteeta} Note that $\eta=1/4$ in the original formulation of the 
maximum force conjecture \cite{Gib,Sch1,Sch2}.
\bibitem{NoteUnits} We shall henceforth use gravitational units with $G=c=1$. 
In these units the radially-dependent force function (\ref{Eq2}) of a curved spacetime is dimensionless.
\bibitem{Noteppd} Here we have used the relation $p(r=r_{\text{max}})>0$ at the maximum 
point $r=r_{\text{max}}$.
\bibitem{Notefp} Here we have used the relation ${\cal F}_{\text{max}}\equiv {\cal F}(r=r_{\text{max}})>0$.
\bibitem{Noterho} Here we have used the relation $\rho\geq p>0$ at the maximum point $r=r_{\text{max}}$.
\bibitem{Notegen} It is worth emphasizing again that the analytically derived bounds (\ref{Eq24}) and (\ref{Eq27}) are 
valid for generic (that is, not necessarily isotropic) field configurations.
\end{thebibliography}
\end{document}